\documentclass[a4paper,11pt]{amsart}
\usepackage{amssymb}
\usepackage{amsmath}
\usepackage{graphicx}

\input{tcilatex}

\begin{document}
\thanks{The project is realized under Marie Curie grant PIEF-GA-2009-253845}
\title{}
\author{}
\maketitle

\begin{center}
{\Large \textbf{The nest site lottery:}}

{\Large \textbf{how selectively neutral density dependent growth }}

{\Large \textbf{suppression}}

{\Large \textbf{induces frequency dependent selection}}\bigskip 

Theoretical Population Biology 90 (2013) 82-90\\[0pt]

\bigskip 

K.~Argasinski* \\[0pt]
Department of Mathematics, University of Sussex, \\[0pt]
Brighton BN1 9QH, UK.

K.Argasinski@sussex.ac.uk, argas1@wp.pl

tel.\ 012 73877345

\bigskip

M.~Broom \\[0pt]
Department of Mathematics, City University London, \\[0pt]
Northampton Square, London EC1V 0HB, UK.

Mark.Broom.1@city.ac.uk\\[0pt]

\bigskip 
\end{center}

*corresponding author

The project is realized under grant Marie Curie grant\ PIEF-GA-2009-253845.

\newpage

\textbf{Abstract}

Modern developments in population dynamics emphasize the role of the
turnover of individuals. In the new approaches stable population size is a
dynamic equilibrium between different mortality and fecundity factors
instead of an arbitrary fixed carrying capacity. The latest replicator
dynamics models assume that regulation of the population size acts through
feedback driven by density dependent juvenile mortality. Here, we consider a
simplified model to extract the properties of this approach. We show that at
the stable population size, the structure of the frequency dependent
evolutionary game emerges. Turnover of individuals induces a lottery
mechanism where for each nest site released by a dead adult individual a
single newborn is drawn from the pool of newborn candidates. This frequency
dependent selection leads toward the strategy maximizing the number of
newborns per adult death. However, multiple strategies can maximize this
value. Among them, the strategy with the greatest mortality (which implies
the greatest instantaneous growth rate) is selected. This result is
important for the discussion about universal fitness measures and which
parameters are maximized by natural selection. This is related to the
fitness measures $R_{0}$ and $r$, because the number of newborns per single
dead individual equals lifetime production of newborn $R_{0}$ in models
without ageing. We thus have a two-stage procedure, instead of a single
fitness measure, which is a combination of $R_{0}$ and $r$. According to the
nest site lottery mechanism, at stable population size, selection favours
strategies with the greatest $r$, i.e. those with the highest turnover, from
those with the greatest $R_{0}$.\bigskip


\section{Introduction}

In the modern theory of evolutionary ecology (Post and Palkovacs 2009,
Pelletier et al. 2009, Morris 2011, Schoener 2011) the problem of
eco-evolutionary feedback is of special interest. One of the major
theoretical problems in the modelling of population dynamics, and in general
of evolutionary biology and ecology, is the limit of population growth and
its selection consequences. This topic is very important in many disciplines
such as evolutionary game theory and life history theory.

The earliest attempt to solve this problem for populations with overlapping
generations is the continuous logistic equation introduced by Verhulst in
the 19th century (Verhulst, 1838), which can be found in every textbook on
ecology and mathematical biology. It inspired the idea of $r$ and $K$
selection (McArthur and Wilson 1967), that selection favours different
strategies at low densities and near the stable population size, and is
still applied in modelling (Cressman et al. 2004, Cressman and Krivan 2006,
Cressman and Krivan 2010). This concept states that there is some arbitrary
maximal population size at which growth is suppressed and the population
remains stable. However, this approach produces some unusual predictions
which provoked a wide discussion (Koz\l owski 1980, \L omnicki 1988, Kuno
1991, Ginzburg 1992, Gabriel 2005, Hui 2006, Argasinski and Koz\l owski
2008) presented in the next section.

The problem of the limits to growth is important not only for ecological
population growth models but also for the modelling of natural selection.
There is huge discussion on what is maximized by natural selection and what
happens when a population reaches the limit to growth (Metz et al. 1992, Koz%
\l owski 1993, Mylius and Diekmann 1995, Brommer 2000, Dieckmann and Metz
2006, Metz et al. 2008, Roff 2008). However in these attempts density
dependence is represented by some abstract unspecified factors. Thus the
proposed solutions are very general and abstract. A concrete mechanistic
interpretation should be helpful in the interpretation of the mathematical
notions. In the modification of the logistic equation (Koz\l owski 1980, Hui
2006) which was applied to game-theoretic modelling (Argasinski and Koz\l %
owski 2008, Argasinski and Broom 2012) there is an example of a mechanism
responsible for strategically neutral density dependence called in this
paper a "nest site lottery". The underlying assumption is that there is a
limited number of nest sites in the environment and that every newborn must
find a nest site to survive. Thus all newborns produced at some moment in
time form a pool of candidates to be drawn from to replace the dead
individuals in their nest sites. The difference is that in this case there
is an arbitrary maximal population size described by a carrying capacity
indicating the number of available nest sites (mechanistically interpreted
as nests or holes, where individual can settle, as in Hui 2006). However,
the stable population size is not the carrying capacity, as in the classical
logistic equation, but the dynamic equilibrium between different factors of
mortality and fecundity (Koz\l owski 1980, Ginzburg 1992, Hui 2006), which
can be affected by the dynamics of the population state (Argasinski and
Broom 2012). The advantage of this approach is that it considers a realistic
turnover of individuals (Argasinski and Koz\l owski 2008). In this paper we
will more rigorously analyze the properties of the nest site lottery
mechanism in a simpler model than in the previous papers (Argasinski and Koz%
\l owski 2008, Argasinski and Broom 2012).

Below we give the mathematical details of this approach (sections 2 and 3).
Section 4 starts the development of the selection model and in section 5 the
important notion of the turnover coefficient is introduced. Section 6
contains the presentation of the selection dynamics. In section 7 the
relationships between the nest site lottery mechanism and the invasion
fitness concept are presented. Section 8 contains the main results which are
the equations (\ref{intra}, \ref{inter}) and their analysis describing the
nest site lottery mechanism (intuitively depicted in fig. 1). We see that,
eventually, selection favours the strategy with the highest turnover
coefficient out of those with the greatest value of lifetime reproduction.
The mechanistic reasoning from section 8 is completed by Theorem 1
describing the quantitative characterization of the restpoints of the system
(\ref{intra}, \ref{inter}). The importance of the obtained results and the
general ideas inspired by them is discussed in section 9 (the last
subsection contains a discussion on the two-stage maximization procedure,
substituting for the single-step fitness measure, obtained by our results).
\bigskip

\section{Density dependence}

The cornerstone of mathematical ecology is the Malthusian equation
describing exponential\ population growth, 
\begin{equation}
\frac{dn}{dt}=nr=nb-nd=n\left( b-d\right) ,  \label{eq:popgrowth}
\end{equation}

where $b$ is the birth rate and $d$ is the death rate. However, in
Argasinski and Broom (2012, see Appendix 1 there for details) it was shown
that with respect to the multiplicative proportionality constant (which can
be removed using a change of timescale) acting as the rate of interaction
occurrence, these parameters can be interpreted as demographic parameters
describing the outcomes of the average interaction with elements of the
environment or other individuals. Then $b\in \lbrack 0,\infty )$ can be
interpreted as the number of newborns produced during an interaction event
and $d\in \lbrack 0,1]$ as the probability of death during an interaction
event. We will assume this mechanistic interpretation in our model. Thus the
Malthusian coefficient $r$ can be interpreted as the balance between
mortality $d$ and fertility $b$. The above model is not realistic, because
it allows for infinite population growth. The classical solution of this
problem is the use of the logistic equation, which is equation (\ref%
{eq:Verhulst}),%
\begin{equation}
\frac{dn}{dt}=nr\left( 1-\dfrac{n}{K}\right) .  \label{eq:Verhulst}
\end{equation}

However, this relies on a problematic assumption which has very serious
consequences. Equation (\ref{eq:Verhulst}) produces artifacts in population
growth models (Kuno 1991, Gabriel 2005) and selection models related to
replicator dynamics (Argasinski and Koz\l owski 2008). For example, it
suppresses the selection dynamics in the replicator dynamics by setting the
right hand sides of the strategy dynamics equations to 0 (Argasinski and Koz%
\l owski 2008), the trajectory escapes to infinity for $r<0$ (i.e. $b<d$)
and initial population size greater than $K$ (known as Levins' paradox,
(Gabriel 2005)) or the trajectory decreases with increasing rate for $r<0$
and initial population size slightly smaller than $K$ (Kuno 1991). This is
caused by the fact that the term $r$ is multiplied by the suppression
coefficient, which implies that with population growth, both mortality and
fertility decrease, and mortality decreasing with increasing population size
and reaching zero at equilibrium is biologically counterintuitive. Mortality
should not decrease with population growth and individuals cannot be
immortal at equilibrium. The above problems suggest that models should rely
on clear and mechanistic assumptions (Geritz and Kisdi 2012). Thus, density
dependent suppression should act only on the number of juveniles recruited
to the population (Koz\l owski 1980, Ginzburg 1992) and the initial
population size should be smaller than the carrying capacity (Hui 2006,
Argasinski and Koz\l owski 2008, Argasinski and Broom 2012) leading to%
\begin{equation}
\dot{n}=n\left( b\left( 1-\dfrac{n}{K}\right) -d\right),
\label{eq:onlyfertility}
\end{equation}

where the suppression term $\left( 1-n/K\right) $ describes newborns'
survival probability. This provides an important distinction between newborn
candidates introduced to the environment (described by per capita number $b$%
) and recruited newborns, survivors of the density dependent stage
(described by $b\left( 1-\dfrac{n}{K}\right) $).

This problem was emphasized by Koz\l owski (1980) for the first time, but
surprisingly this paper did not get as wide an appreciation as it deserved.
This problem was also mentioned in the classical book "Population Ecology of
Individuals" (\L omnicki 1988). Then it was reinvented by Ginzburg (1992),
but (\ref{eq:onlyfertility}) was rejected there as it \textquotedblleft
disagrees with our intuition about unchanging equilibrium \textquotedblright
. Hui (2006) argued, against Ginsburg's claim, that (\ref{eq:onlyfertility})
is the proper approach and should be substituted for (\ref{eq:Verhulst}).
The discussion started by Ginzburg also did not receive wide attention.
Argasinski and Koz\l owski (2008) then applied equation (\ref%
{eq:onlyfertility}) to avoid the suppression of selection that occurs after
the equilibrium size is reached caused by equation (\ref{eq:Verhulst})
without knowledge of this discussion, and (\ref{eq:onlyfertility}) is a
cornerstone of the ecologically realistic approach to dynamic evolutionary
games (Argasinski and Broom 2012). Then (\ref{eq:onlyfertility}) was
mentioned as an example of the proper mechanistic approach (Geritz and Kisdi
2012), but not as the general alternative to (\ref{eq:Verhulst}). However,
we believe that (\ref{eq:onlyfertility}) deserves much stronger attention
from a general audience.

Although (\ref{eq:onlyfertility}) has been applied in complex selection
models (Argasinski and Koz\l owski 2008, Zhang and Hui 2011, Argasinski and
Broom 2012), the selection consequences of this approach have not been
rigorously analyzed, since previous papers (Kuno 1991, Ginzburg 1992,
Gabriel 2005, Hui 2006) focused on population density dynamics and
ecological aspects. This distinction between adults and newborn candidates
is very important for ecological and evolutionary reasoning, because
differences between juvenile and adult mortality can have serious selection
consequences. For example, a lack of mortality differences means that a
small fecundity advantage can favour evolution of semelparity over
iteroparity (this problem is known as Coles Paradox, Cole 1954), while
mortality differences can significantly change the situation (Charnov and
Shaffer 1973). The selection mechanism induced by (\ref{eq:onlyfertility})
is thus very interesting and will be analyzed in later sections.

\section{The population in equilibrium}

We can calculate equilibrium size, by setting the right hand side of
Equation (\ref{eq:onlyfertility}) to be equal to 0, which gives either 
$n=0$ or 
\begin{equation}
\tilde{n}=\left( 1-\dfrac{d}{b}\right) K.  \label{eq:equilibrium}
\end{equation}

Note that for positive $\tilde{n}$, the condition $b>d$ should be satisfied.
After substitution of $\tilde{n}$ into the logistic coefficient $(1-n/K)$,
we obtain the equilibrium newborn survival probability $d/b$. This is
reasonable; due to the turnover of individuals, in any short time interval
for every $nb$ newborns we have $nd$ dead individuals. Thus $nd/nb$
describes the number of newborns competing for each single nest site vacated
by a dead individual. Only one newborn can settle in a single place, thus
each newborn can survive with probability $d/b$. This newborn survival
should be valid for any density dependent mortality acting on juveniles, not
only for logistic suppression, because only in this case does fertility
equal overall mortality.

\section{The case of multiple individual strategies}

Assume that there are different individual phenotypes $i=1,\ldots ,H$ each
characterized by per capita reproduction $b_{i}$ and mortality $d_{i}$. Thus
every strategy is described by a two dimensional vector $v_{i}=[b_{i},d_{i}]%
\in \left( \lbrack 0,\infty )\times \lbrack 0,1]\right) $ describing
demographic parameters interpreted as in (\ref{eq:popgrowth}). Note that $K$
describes the number of nest sites and is the same for all phenotypes.
Denoting $n=\sum_{i}n_{i}$ and $q_{i}=\dfrac{n_{i}}{n}$, we can describe the
following dynamics:%
\begin{equation}
\frac{dn_{i}}{dt}=n_{i}\left( b_{i}\left( 1-\dfrac{n}{K}\right)
-d_{i}\right) .  \label{eq:nidynamics}
\end{equation}

The value $n_{i}$ increases with time if 
\begin{equation}
n<\left( 1-\dfrac{d_{i}}{b_{i}}\right) K  \label{ncritical}
\end{equation}%
and decreases in the opposite case. Thus for every strategy there is a
critical population size which is a threshold between regions of growth and
decline. Above the population size critical for a particular strategy, the
effective fertility $b_{i}\left( 1-\dfrac{n}{K}\right) $ will be smaller
than the mortality $d_{i}.$\ Thus\ the dynamics of the population size plays
an important role, which is described by the equation%
\begin{eqnarray*}
\frac{dn}{dt} &=&\sum_{i}\dot{n}_{i}=\sum_{i}n_{i}\left( b_{i}\left( 1-%
\dfrac{n}{K}\right) -d_{i}\right) \\
&=&n\left( \left( 1-\dfrac{n}{K}\right)
\sum_{i}q_{i}b_{i}-\sum_{i}q_{i}d_{i}\right) ,
\end{eqnarray*}%
giving 
\begin{equation}
\frac{dn}{dt}=n\left( \left( 1-\dfrac{n}{K}\right) \bar{b}-\bar{d}\right) ,
\label{eq:ndynamics}
\end{equation}%
where $\bar{b}(q)=\sum_{i}q_{i}b_{i}$ and $\bar{d}(q)=\sum_{i}q_{i}d_{i}$.
We can easily calculate that in this case, instead of reaching the stable
equilibrium, the population size converges to the stationary density
manifold (Cressman et al 2001,Cressman and Garay 2003a and b)%
\begin{equation}
\tilde{n}=\left( 1-\dfrac{\bar{d}}{\bar{b}}\right) K,  \label{stablepopsize}
\end{equation}

the form of which is conditional on the strategy frequencies. Thus we
introduced diversity among individual strategies to our model.\ In our
model, in the general case, newborn survival $\left( 1-\dfrac{n}{K}\right) $
is a phenomenological function, linear with respect to the fraction of free
nest sites. Thus in this approach the recruitment probability equals the
probability of finding a free nest site in a single trial. This is a very
specific mechanism which will not be suitable for many species. However,
similar mechanisms will work for any density dependent factor $u(n)$ acting
on births that is monotonically decreasing with respect to $n$. Then for
growth rate $b_{i}u(n)-d_{i},$ the critical population size will be $%
n=u^{-1}\left( \dfrac{d_{i}}{b_{i}}\right) .$\ The newly produced offspring
of the carriers of the different strategies form a pool of candidates from
which randomly drawn individuals will be recruited to settle in the
available nest sites. This is the core of the "nest site lottery" mechanism
which will be analyzed in the following sections. Note that equations (\ref%
{eq:nidynamics}) and (\ref{eq:ndynamics}) suggest the importance of the
factors $\dfrac{d_{i}}{b_{i}}$ and $\dfrac{\bar{d}}{\bar{b}}$. This will be
analyzed in the next section.

\section{The turnover coefficient L}

Here we will introduce an important characterization of population dynamics.
We shall define the function $L(v)=b/d$ for a single strategy $v$. $L$
describes the number of newborns per single dead individual, which we shall
refer to as the\ turnover coefficient (for the relationship of the turnover
coefficient with lifetime reproduction, see the Discussion). Surprisingly, a
similar coefficient describing the energy allocated to reproduction divided
by mortality can be found in life history papers (Taylor and Williams 1984,
Koz\l owski 1992, Koz\l owski 1996, Werner and Anholt 1993, Perrin and Sibly
1993, for an overview see Koz\l owski 2006). Analogously, for a mixture of
strategies where $\bar{v}(q)=\sum_{i}q_{i}v_{i}=[\bar{b},\bar{d}]$ is the
average strategy contained in the convex hull of the strategies $v_{i}$ (see
fig. 1), we define $L(\bar{v}(q))=\bar{b}/\bar{d}$. Thus 
\begin{equation}
L(\bar{v}(q))=\frac{\bar{b}}{\bar{d}}=\frac{\sum_{i}q_{i}b_{i}}{%
\sum_{i}q_{i}d_{i}}=\frac{\sum_{i}q_{i}d_{i}L(v_{i})}{\sum_{i}q_{i}d_{i}}%
=\sum_{i}\frac{q_{i}d_{i}}{\sum_{j}q_{j}d_{j}}L(v_{i})=\sum_{i}y_{i}L(v_{i}),
\end{equation}%
which is a weighted average of the $L(v_{i})$s and $y_{i}=q_{i}d_{i}/%
\sum_{j}q_{j}d_{j}$ describes the fraction of $i$ strategists among
individuals dying during a small time interval $\Delta t$ (according to
Appendix A, \ref{subpop}). $L(\bar{v}(q))$ is thus the average $L$ among
dead adult individuals. The $L$-function can be useful in describing the
multiplicative newborn survival (recruitment probability) because after
substitution of the stable population size $\tilde{n}$ into the logistic
suppression coefficient we obtain: 
\begin{equation}
\left( 1-\dfrac{\tilde{n}}{K}\right) =\bar{d}/\bar{b},  \label{eq:newborn}
\end{equation}%
which can be denoted as $1/L(\bar{v}(q))$.

If there is any variation in the $L(v_{i})$s, then we have that $\bar{b}/%
\bar{d}$ lies strictly between the smallest and largest values of $L$, $%
L_{\min }<\bar{b}/\bar{d}<L_{\max }$. $L(v)$ describes the number of newborn
candidates produced per single dead individual for the strategy $v=\left[ b,d%
\right] $, during $\Delta t$. When the strategic argument is the averaged
vector, describing a population with a mixture of strategies, then the value
of $L$ is the average number of newborn candidates produced per single dead
individual in this population. When the population\ is in size equilibrium
(at the stationary density manifold), then the newborn survival component
can be described by the value of $L$ of the average population strategy; \
thus it becomes frequency dependent.

\section{Selection dynamics}

The behaviour of equation (\ref{eq:nidynamics}) suggests frequency dependent
self-regulation of the population state. Equation (\ref{eq:ndynamics}) shows
attraction to the stable size manifold (\ref{stablepopsize}), which suggests
that the dynamics on this manifold should be analyzed. To describe the
frequency dependent selection associated with the system we have presented,
tools appropriate to game dynamics are required. Thus we should describe the
population in terms of the strategy frequencies $q_{i}$ and the population
size $n$. However, at the stable size manifold the population size is given
by (\ref{eq:newborn}). We can assume that the strategies are close enough to
each other that a separation of timescales between fast $n$ dynamics and $q$
dynamics occurs. Then we can assume that selection occurs on the stationary
size manifold. Now we can describe the selection process realized by the
"nest site lottery" mechanism. Thus using (\ref{eq:newborn}), we can write
the selection dynamics from (\ref{eq:nidynamics}) as 
\begin{equation}
\frac{dn_{i}}{dt}=n_{i}\left( b_{i}\dfrac{\bar{d}(q)}{\bar{b}(q)}%
-d_{i}\right) .  \label{growth}
\end{equation}

Because the average growth rate on the stable size manifold is zero then the
equation (\ref{growth}) can be replaced by the replicator dynamics (see
Appendix A) 
\begin{equation}
\frac{dq_{i}}{dt}=q_{i}\left( b_{i}\dfrac{\bar{d}(q)}{\bar{b}(q)}%
-d_{i}\right) =q_{i}d_{i}\left( \dfrac{L(v_{i})}{L(\bar{v}(q))}-1\right) .
\label{rep}
\end{equation}

Therefore the growth rate of the $i$-th\ strategy becomes a\ function of the
strategy frequencies $q$\ (frequency dependent):%
\begin{equation*}
M(v_{i},q)=b_{i}\dfrac{\bar{d}\left( q\right) }{\bar{b}\left( q\right) }%
-d_{i}=d_{i}\left( \dfrac{L(v_{i})}{L(\bar{v}(q))}-1\right) ,
\end{equation*}%
and by equations (\ref{eq:ndynamics}) and (\ref{stablepopsize}) the
stationary population size\ manifold is described by 
\begin{equation*}
\tilde{n}=\left( 1-\dfrac{\bar{d}\left( q\right) }{\bar{b}\left( q\right) }%
\right) K=\left( 1-\dfrac{1}{L(\bar{v}(q))}\right) K.
\end{equation*}

Note that the growth rate function $M$ describes a mixture of all mortality
and fecundity components, not only the density independent mortality $d$ and
fecundity $b$ as in the Malthusian parameter $r$. The growth rate is
positive when $L(v_{i})>L(\bar{v}(q))$, which implies that 
\begin{equation}
\dfrac{b_{i}}{d_{i}}>\dfrac{\sum_{j}q_{j}b_{j}}{\sum_{j}q_{j}d_{j}}.
\label{threshold}
\end{equation}

Thus there is a threshold between regions of growth (strategies with
reproductive surplus) and reduction (strategies with death rate exceeding
birth rate) which has the linear form $b_{i}=L(\sum_{j}q_{j}v_{j})d_{i}$
(see Figure 1). The threshold describes the set of strategies for which the
growth rate $M(v_{i},q)$ equals 0. Frequencies $q_{i}$ of strategies $v_{i}$
with a greater value of $L$ than the average strategy $\sum_{j}q_{j}v_{j}$
will increase under the replicator dynamics. In effect the averaged strategy
shifts towards those strategies because it is a linear combination of the
strategies present in the population. This implies an increase of $L$ of the
average strategy (see Figure 1). However, among growing strategies, the
greatest growth rate is by the strategy with the greatest coefficient $%
M(v_{i},q)$. Because in this case the dynamics is on the stationary density
manifold, the current population size is very close to (\ref{stablepopsize})
and \ (\ref{threshold}) is equivalent to satisfying inequality (\ref%
{ncritical}) (passing the critical population size). Thus at the stationary
size manifold the threshold between the growth and decline of the strategy
frequency is equivalent to the threshold between the growth and decline of
the number of carriers of that strategy (this may not be satisfied far from
the stationary size manifold). Frequency dependence induces an increase of
the slope of the threshold which eventually leads to the selection of the
strategy with the greatest $L$, which confirms the result of Mylius and
Diekmann (1995). Note that their second result, that density dependent adult
mortality leads simply to $r$ maximization as in unlimited growth models,
directly comes from the independence of the replicator dynamics from
background fitness.

\section{The monomorphic resident-mutant case}

We can simplify the above model by assuming a monomorphic population invaded
by a rare mutant; thus this resembles the classical ESS approach (Maynard
Smith 1982) in the context of life history evolution (Charlesworth \& Leon
1976, Mylius and Diekmann 1995). In the limiting case where the strategy
trait tends to zero, we approach the method known as invasion analysis which
is the cornerstone of adaptive dynamics (Dieckmann and Law 1996, Metz et al.
1996, Geritz et al. 1998, Dercole and Rinaldi 2008). Using Equation (\ref%
{rep}), the resident growth rate is zero and the rare mutant growth rate
function $M$ is $b_{mut}(d_{res}/b_{res})-d_{mut}$ which must be positive to
invade the population. Thus the equilibrium population size increases.

To be an ESS itself, the \textquotedblleft mutant\textquotedblright\
population should be stable against the previous resident, and thus: 
\begin{equation*}
b_{res}\dfrac{d_{mut}}{b_{mut}}-d_{res}<0\Rightarrow \dfrac{b_{res}}{d_{res}}%
<\dfrac{b_{mut}}{d_{mut}},
\end{equation*}%
which is the same condition. Thus in both cases we obtain $L_{res}<L_{mut}.$

Note that, when we consider only the death component of the Malthusian
equation then we obtain the equation $\dot{n}=-nd$; thus this is an
exponential decay with decay constant $d$, and so the average lifetime of
the individual is $\kappa =1/d$. In any short time interval of length $%
\Delta t$, for every $nb$ newborns we have $nd$ dead individuals. We can
change the timescale to set $\Delta t$ as the new time unit. Then initial
rates, and thus respective births and deaths numbers, should be multiplied
by some timescale specific constant. However this constant cancels out in $%
L(v)$. Thus $L(v)$ is the lifetime reproduction $R_{0}$. Therefore we have
obtained for the "nest site lottery" mechanism, the classical result that
under limited growth only lifetime reproduction is maximized and there is no
selection pressure on the lifespan. However this occurs only in a
monomorphic resident-mutant model. The case of a population composed of an
arbitrary number of individual strategies is more interesting. \bigskip

\section{Multiple strategies with $L=L_{\max }$}

We have seen that evolution leads to the fixation of the strategy with the
largest value of $L$, $L_{\max }$. What if there is more than one such
strategy? The following question arises: is there selection between
strategies with the same $L$? We can show this by applying a multipopulation
game-theoretic approach (Appendix A and Argasinski 2006) and divide
strategies present in the population among subpopulations with the same $L$,
but different $d$s (Appendix B). Assume than we have $m$ such classes with $%
H_{j}$ different strategies in the $j$-th $L$-class (then the lower strategy
index descxribes the number within the particular $L$-class and the upper
strategy index describes the $L$-class). Then for all strategies (for all $i$%
) from the same $L$-class $L(v_{i}^{j})=L^{j}$. When we assume that the
dynamics is on the stable density manifold, we obtain the following
equations 
\begin{eqnarray}
\frac{dq_{i}^{j}}{dt} &=&q_{i}^{j}\left( \dfrac{L^{j}}{L(\bar{v}(q))}%
-1\right) \left( d_{i}^{j}-\sum_{w}q_{w}^{j}d_{w}^{j}\right) ,  \label{intra}
\\
\frac{dg_{j}}{dt} &=&g_{j}\left( \dfrac{L^{j}}{L(\bar{v}(q))}-1\right)
\sum_{w}q_{w}^{j}d_{w}^{j},  \label{inter}
\end{eqnarray}

describing the changes of the proportion of the $i$-th strategy within the $%
j $-th $L$-class (Equation \ref{intra}) described by $q_{i}^{j}$ and related
frequencies between $L$-classes (\ref{inter}) described by $g_{j}$. Thus
selection between $L$-classes is driven by the first bracketed term from
Equation (\ref{intra}) and affects both intra- and inter-group dynamics.
However, there is selection inside each $L$-class toward greater $d$. When
suboptimal $L$-classes are outcompeted, the intrinsic selection driven by
the bracket $\left( d_{i}^{j}-\sum q_{w}^{j}d_{w}^{j}\right) $ is also
suppressed. The form of Equation (\ref{inter}) shows that among growing $L$%
-classes, those with smaller $L$ can grow faster than those with larger, due
to a greater $\sum q_{w}^{j}d_{w}^{j} $, until they fall under the $L$%
-selection threshold.

We are interested in analyzing which strategy (strategies) will dominate the
population in the long term. In particular, $\dot{q}_{i}$ in Equation (\ref%
{rep})\ is always positive if strategy $i$ has $L(v_{i})=L_{\max }$,
whenever there is variation in the $L$ values in the population. Thus the
proportions of such strategies increase; but also, following Equation (\ref%
{intra}), the strategies out of these with the largest values of $d_{i}$
increase the fastest. Thus if there is either repeated small mutations
involving strategies with $L(v_{i})<L_{\max }$ or a constant low level of
mutation involving a mix of strategies making $L(\bar{v}(q))<L_{\max }$, the
population will evolve to the strategy out of those with $L(v_{i})=L_{\max }$
such that $d_{i}$ takes the largest value. Thus repeated mutations or
invasions of suboptimal strategies induce selection towards maximal $d$
among $L_{\max }$ strategists. It is easy to show that in the absence of
density dependent suppression, this strategy has the greatest $r$ but only\
among $L_{\max }$ strategies, since $b_{i}=d_{i}L_{\max }$. The strategies
from other $L$-classes can have even greater $r$, but they will be
outcompeted by the mechanism described by the bracketed term in equation (%
\ref{rep}). In the case when the population consists only of the $L_{\max }$
individuals, the same outcome can be caused by repeated ecological
catastrophes leading to a decrease of the population size. Then the strategy
with the greatest $d_{i}$ will have the greatest growth rate during the
growth phase of the population.

However, let us focus on the evolution of the system under the replicator
dynamics in a single particular \textquotedblleft turn\textquotedblright ,
during which no mutation occurs. Suppose that there are precisely $I$
strategies in the $L_{\max }$-class, and assume that the initial state of
the $L_{\max }$-class is described by vector $q=(q_{1}^{\max }(0),\ldots
,q_{I}^{\max }(0))$\ and initial relative size $g_{\max }$. \bigskip

\textbf{Theorem 1} \newline
The replicator dynamics converges to the vector \newline
$q=(q_{1}^{\max }(0)g_{\max }(0)\lambda ^{d_{1}},\ldots ,q_{I}^{\max
}(0)g_{\max }(0)\lambda ^{d_{I}})$, \newline
where $\lambda $ is a constant that satisfies the equation $g_{\max
}(0)\sum_{i=1}^{I}q_{i}(0)\lambda ^{d_{i}}=1$.

For the proof see Appendix C. \bigskip

Theorem 1 shows that the restpoint describing the frequencies among $L_{\max
}$ strategists is fully determined by the initial state of the $L_{\max }$%
-class and\ its initial relative size $q_{1}^{\max }(0)$. Despite frequency
dependence, this occurs independently of the initial frequencies of the
other strategies (note that in our model there are no direct interactions
between individuals). Thus calculation of $q$ reduces to the finding of the
appropriate value of $\lambda $. Note that the rest point $q$\ can be
interpreted as the state of the whole population in general coordinates and
the final state of the $L_{\max }$-class (then $g_{\max }=1$). Note that if
initially $g_{\max }(0)=1$, then obviously $\lambda =1$. The parameter $%
\lambda $\ can be described as the inflation coefficient because it inflates
the frequencies to sum them to one and compensate the impact of $g_{\max
}(0) $.\bigskip

\section{Discussion}

\subsection{How the nest site lottery works?}

We started from the basic population growth equation which is the
cornerstone of the framework underlying evolutionary game theory and
replicator dynamics (Maynard Smith 1982, Cressman 1992, Hofbauer and Sigmund
1988 and 1998) and its more ecologically realistic extensions (Cressman and
Garay 2003a,b, Argasinski 2006, Argasinski and Koz\l owski 2008, Argasinski
and Broom 2012). We presented an analysis of the dynamics of the mechanism
inducing frequency dependent selection toward the strategy maximizing the
turnover coefficient $L(v_{i})$.

This phenomenon can be explained mechanistically. All newborns introduced
into the population at the same moment in time form a pool of candidates.
Each newborn has equal probability to survive (find a nest place), thus the
strategy maximizing the number of newborns (trials) maximizes the fraction
in the pool of candidates and in effect the amount of survivors. However,
every dead adult can be substituted by an individual with any other
strategy, thus each death is an additional free place in the lottery. Thus
it is profitable for the strategy carried by some subpopulation to maximize
the number of trials (newborns) per single offered place (dead adult). In
addition, we have shown that among strategies with the largest value of the
turnover coefficient $L_{\max }$ there is a selection pressure toward the
strategy with the greatest $d$. This is intuitive from equation (\ref{rep}),
because for strategies with the maximal number of newborn candidates
produced per dead adult (i.e. maximizing the bracketed term in equation \ref%
{rep}), the growth rate will increase with the number of dead adults
(described by the fraction $q_{i}d_{i}$ in equation \ref{rep}) since each of
them will be exchanged for $L_{\max }$ newborns in the pool of candidates.
Note that this provides a gene centered mechanistic explanation of the
phenomenon which can be naively interpreted in terms of group selection and
an altruistic "sacrifice" of adults, to release the nest sites for
juveniles.However, our model shows that it is an outcome of "selfish"
fitness maximization at the individual level. In addition our model suggests
a possible tradeoff in resource allocation between maximization of the
number of candidates in the nest site lottery and survival of the parental
individual.

\subsection{Importance of the nest site lottery mechanism}

The model presented in this paper is as simple as possible, to emphasize the
mechanistic aspects of the analyzed phenomenon. For example, there are no
direct interactions between individuals as in game-theoretic models. Our
model has an extremely simplified age structure consisting only of juveniles
and adults (more on the limitations of pure age-dependent models can be
found in Metz and Diekmann 1986). However, it was shown that the\ impact of
density dependent factors (thus\ also the mechanism described in this paper
and its generalizations) can significantly affect and alter the outcomes of
game-theoretic models (Argasinski and Koz\l owski 2008, Argasinski and Broom
2012). This is caused by the feedback driven by the fact that when
population size is on the stable size manifold every newborn should find a
new nest site vacated by a dead adult. Our model is the simple case example
of a one-dimensional monotone density dependence acting on the effective
birth rate (more on this and other general cases can be found in Metz et al
2008). However, as was shown at the end of section 4, our results can be
extrapolated to other factors that are monotonically decreasing with respect
to the population size, acting like juvenile survival. 
It is possible that other environmental feedback loops of the same type to
those in our model may induce similar selection mechanisms.

The mechanism shown in this paper supports the intuition underlying $r$ and $%
K$ selection theory (McArthur and Wilson 1967), that natural selection
favours different strategies in growing populations than in populations with
suppressed growth. The theoretical and methodological aspects of this
approach were criticized (Barbault 1987, Getz 1993, Stearns 1977), however
as an intuition it still seems to be relevant\ (for modern approaches see
for example Metz et al. 2008). An alternative to the $r$ and $K$ approach is
life history theory (Roff 1992, Stearns\ 1992), where the problem of
different selection mechanisms in limited and unlimited populations also
exists. Maybe phenomena similar to those revealed by our simple model can be
found in other, more general or different specific models related to general
population dynamics, life history evolution, adaptive dynamics or population
genetics. This can be the subject of future research.

\subsection{What is maximized by natural selection, and when?}

The exact meaning of \textquotedblleft fitness\textquotedblright\ is a
subject of endless discussion (Metz et al. 1992, Koz\l owski 1993, Mylius
and Diekmann 1995, Brommer 2000, Dieckmann and Metz 2006, Metz et al. 2008,
Roff 2008). Basically "fitness" can be defined as the instantaneous growth
rate or invasion exponent (Metz 2008). However, if eco-evolutionary feedback
is of a particulary simple kind, the optimization approach can be applied
(Metz et al 2008, Gyllenberg et al 2011) where some "fitness measures" or
"proxies" are maximized. There is a widely known fact in life history theory
that in a population with unlimited growth, the Malthusian growth rate $r$
is a proper fitness measure, while on the stable size manifold, lifetime
production of newborns (before juvenile mortality selection) $R_{0}$ is the
correct measure. In Mylius and Diekmann (1995) there is a statement that the
invasion fitness method (Metz et al. 1992) suggests that $R_{0}$ and $r$ are
necessarily both maximal at the ESS (although this statement is unclear
since some strategies can maximize $R_{0}$ and others can maximize $r$). Our
results support this claim, and show that an analogous mechanism can act in
population dynamic and game-theoretic models.

Here an important claim is that of Brommer and Kokko (2002), who say that $%
R_{0}$ is a rate independent reproductive measure which does not depend on
the timing of reproductive events. This is because $R_{0}$ is described on
the lifespan timescale, not the population dynamic timescale like $r$.
Despite its simplified form our model can be a useful illustrative example
for this problem. How can we use a lifespan perspective in our approach? At
first, assume that population growth is unlimited (every newborn candidate
can find a nest site). Then demographic parameters $b$\ and $d$ describing
strategy $v$,\ are constant and the average lifetime of the individual is $%
\kappa (v)=1/d$ (as in section 7). Then $r(v)=b-d=\left( L(v)-1\right)
/\kappa (v)$ and $L(v)$ is the lifetime reproduction $R_{0}$ (or the average 
$R_{0}$ among individuals dying during $\Delta t$ in a population described
by $v$). Therefore, the formula $r=\left( L-1\right) /\kappa $\ shows how
the growth rate is affected by lifetime reproduction in the case of a
non-age structured population with unlimited growth. It shows that an
individual should basically replace itself, but for the strategy growth rate
to be positive requires a reproductive surplus during a lifetime. Now
introduce the limitation of the nest sites causing density dependent
selection. Our results show that under density dependence, the growth rates
of the strategies are affected by the frequency dependent surplus reducing
mechanism, described by newborn survival $1/L(\bar{v})$ which is the
function of the other strategies present in the population. In effect $r(v)$
is replaced by density dependent growth rate $M(v,q)=\left( L(v)/L(\bar{v}%
)-1\right) /\kappa (v)$\ and $R_{0}=L(v)/L(\bar{v})$. Thus the strategies
maximizing the lifetime production of newborn candidates $L(v)$ will
maximize $R_{0}$\ and the bracketed term of $r(v)$. However among $L(v)$
maximizers, the strategy with smallest $\kappa (v)$ will have greatest $r(v)$%
.

\subsection{Conclusion}

Our simple model suggests an insight into the mechanistic nature of
selection under limited growth and has serious interpretational
consequences. It clearly shows that this problem should not be formulated as
the alternative: evolution maximizes $r$ OR $R_{0}$. In our simple model,
when the population reaches a stable size manifold, then a mechanism that
modifies the $r$'s of competing strategies, which are no longer constants,
emerges to select the strategy with maximal $R_{0}$, or with maximal $r$,
among multiple strategies with maximal $R_{0}$. Thus, our model suggests the
existence of another fitness measure which is the combination of $R_{0}$ and 
$r,$ if our reasoning holds in age structured and other more complex models.
However, it will be not a function which should be maximized, but a two
staged procedure. The first stage should identify the strategies maximizing
the turnover coefficient, while the second stage should find strategies with
the greatest $r$ from strategies chosen in the first stage. We note that our
analysis is a simplification, and whereas $R_{0}=L(v_{i})$ in the models
without age structure as presented in this paper, this is not necessarily
satisfied in age structured models. This should be the subject of future
research.\bigskip

\textbf{Acknowledgments:}


The project is realized under grant Marie Curie grant\ PIEF-GA-2009-253845.
We want to thank Jan Koz\l owski, John McNamara, and Franjo Weissing for
their support of the project and helpful suggestions. In addition we want to
thank two anonymous reviewers for detailed comments and helpful
suggestions.\bigskip

\section*{Appendix A Multipopulation replicator dynamics.}

Assume that we have $H$ individual strategies. Standard replicator dynamics
can be derived by rescaling the growth equation $\frac{dn_{i}}{dt}%
=n_{i}M_{i} $ to the related frequencies $q_{i}=n_{i}/\sum_{j}n_{j}$ which
leads to the equation $\frac{dq_{i}}{dt}=q_{i}\left[ M_{i}-\bar{M}\right] $
(where $\bar{M}=\sum_{j}q_{j}M_{j}$). This equation describes the evolution
of strategy frequencies in the unstructured population. However, we might be
interested in the modelling of the structured population divided into
subpopulations such as different sexes, species etc. Assume that we want to
decompose an entire population into $z$ subpopulations.\ Define%
\begin{equation}
k^{j}=[k_{1}^{j},...,k_{H_{j}}^{j}]  \tag{A.1}
\end{equation}

as a vector of indices of strategies exhibited by individuals from the $j$%
-th subpopulation ($k_{i}^{j}\in \{1,...,H\}$, and $H_{j}$ is the number of
strategies in the $j$-th subpopulation). For example the notation $%
k^{2}=[1,3,5]$ means that, in the second subpopulation there are (only)
individuals with strategies $1,3$ and$\ 5$. Every strategy should belong to
a single unique subpopulation. Then according to Argasinski (2006), by the
following change of coordinates%
\begin{equation}
q^{j}=[q_{1}^{j},...,q_{H_{i}}^{j}]=\left[ \dfrac{q_{k_{1}^{j}}}{%
\sum_{i=1}^{H_{j}}q_{k_{i}^{j}}},...,\dfrac{q_{k_{H_{j}}^{j}}}{%
\sum_{i=1}^{H_{j}}q_{k_{i}^{j}}}\right] \ \ \ \ j=1,...,z  \tag{A.2}
\label{subpop}
\end{equation}

we obtain a distribution of relative frequencies of strategies in the $j$-th
subpopulation. The distribution of proportions between subpopulations has
the form%
\begin{equation}
g=[g_{1},...,g_{z}]=\left[ \sum_{i=1}^{H_{1}}q_{k_{i}^{1}},...,%
\sum_{i=1}^{H_{z}}q_{k_{i}^{z}}\right] ,  \tag{A.3}
\end{equation}

where$\ g_{j}$ is the proportion of the $j$-th subpopulation. Every
decomposition into subpopulations can be reduced again to a single
population model by the opposite change of coordinates $q(g,q^{1},...,q^{z})$
where

\begin{equation}
q_{k_{i}^{j}}=g_{j}q_{i}^{j}.  \tag{A.4}  \label{opposite}
\end{equation}

When we apply the above transformations to the replicator equations, we
obtain a set of equations that describes the dynamics inside the
subpopulations (intraspecific dynamics). When the set of strategies in each
subpopulation is characterized by the vector of indices $k^{j}$, then the
system of replicator equations will be:%
\begin{equation}
\frac{dq_{i}^{j}}{dt}=q_{i}^{j}\left[ M_{i}^{j}-\bar{M}^{j}\right] \hspace{%
0.5cm}i=1,...,H_{j}-1,\hspace{0.2cm}j=1,...,z  \tag{A.5}  \label{multiintra}
\end{equation}

\begin{equation}
\frac{dg_{s}}{dt}=g_{s}\left[ \bar{M}^{s}-\bar{M}\right] \hspace{1cm}%
s=1,...,z-1  \tag{A.6}  \label{multiinter}
\end{equation}

where $\bar{M}^{s}=\sum_{i=1}^{H_{s}}q_{i}^{s}M_{i}^{s}$ is the mean fitness
in the $s$-th subpopulation and $\bar{M}=\sum_{s=1}^{z}g_{s}\bar{M}^{s}$. In
practical applications of this method to the modelling of biological
problems, the replicator equations can be defined on the decomposed
population. This will simplify the formulation of the model, because when
strategies are initially assigned to subpopulations, there is then no need
to change their indices. The choice of subpopulations is arbitrary and
depends on the biological assumptions underlying the analyzed problem. For
example, the entire population may be divided into two competing
subpopulations of hosts and parasites or prey and predators. On the other
hand, it may be divided into two subpopulations of males and females, when
interspecific dynamics will describe the evolution of the secondary sex
ratio, and intraspecific dynamics will describe changes of the frequencies
of strategies inside the male and female subpopulations. The subpopulations
can be divided into subsubpopulations, and the entire population may be
transformed into a complex multilevel cluster structure. However, all these
structures are equivalent to a single population replicator dynamics
model.\bigskip

\section*{Appendix B Derivation of equations (\protect\ref{intra}) and (%
\protect\ref{inter}) describing selection strategies inside $L$-classes and
change of sizes of $L$-classes}

Let us assume than we have $m$ such classes with $H_{j}$ different
strategies in the $j$-th $L$-class. In addition, assume that the dynamics is
on the stable size manifold. Then the initial system of the replicator
equations can be transformed into two sets of differential equations.
Firstly, the within $L$-class dynamics (according to \ref{multiintra}):%
\begin{equation}
\frac{dq_{i}^{j}}{dt}=q_{i}^{j}\left( M(v_{i}^{j})-\bar{M}^{j}\right) , 
\tag{B.1}  \label{multi1}
\end{equation}

where $q_{i}^{j}$ is the proportion of the $i$-th strategy in the $j$-th $L$%
-class and $\bar{M}^{j}=\sum_{w}q_{w}^{j}M(v_{w}^{j})=%
\sum_{w}q_{w}^{j}d_{w}^{j}\left( L(v_{w}^{j})/L(\bar{v}(q))-1\right) $.
Secondly, the between $L$-class dynamics (according to \ref{multiinter}):%
\begin{equation}
\frac{dg^{j}}{dt}=g^{j}\left( \bar{M}^{j}-\bar{M}\right) ,  \tag{B.2}
\label{multi2}
\end{equation}

where $g^{j}$ is the proportion of the $j$-th $L$-class and $\bar{M}=0$,
since the population is on the stable size manifold. Since for all
strategies (for all $i$) from the same $L$-class $L(v_{i}^{j})=L^{j}$, after
substitution of the respective formulae into Equations (\ref{multi1}) and (%
\ref{multi2}), we obtain the equations (\ref{intra}) and (\ref{inter}): 
\begin{eqnarray}
\frac{dq_{i}^{j}}{dt} &=&q_{i}^{j}\left( d_{w}^{j}\left( \dfrac{L(v_{i}^{j})%
}{L(\bar{v}(q))}-1\right) -\sum_{w}q_{w}^{j}d_{w}^{j}\left( \dfrac{%
L(v_{w}^{j})}{L(\bar{v}(q))}-1\right) \right)  \TCItag{B.3}  \label{dhdh} \\
&=&q_{i}^{j}\left( \dfrac{L^{j}}{L(\bar{v}(q))}-1\right) \left(
d_{i}^{j}-\sum_{w}q_{w}^{j}d_{w}^{j}\right) ,  \TCItag{B.4} \\
\frac{dg_{j}}{dt} &=&g_{j}\left( \dfrac{L^{j}}{L(\bar{v}(q))}-1\right)
\sum_{w}q_{w}^{j}d_{w}^{j}.  \TCItag{B.5}  \label{sdsd}
\end{eqnarray}

\section*{Appendix C: Proof of Theorem 1}

From (\ref{rep}) we have that 
\begin{equation*}
\frac{dq_{i}}{dt}=d_{i}q_{i}\left( \frac{L(v_{i})}{L(\bar{v}(q))}-1\right)
\end{equation*}%
and so 
\begin{equation}
\left( \frac{L(v_{i})}{L(\bar{v}(q))}-1\right) =\frac{1}{d_{i}q_{i}}\frac{%
dq_{i}}{dt}.  \tag{C.1}  \label{eq:firsteq}
\end{equation}

Consider any pair of strategies $v_{i}=[b_{i},d_{i}]$ and $%
v_{j}=[b_{j},d_{j}]$ from the same $L$-class (i.e. $L\left( v_{i}\right)
=L\left( v_{j}\right) $). Using (\ref{eq:firsteq}) we obtain 
\begin{equation*}
\frac{1}{d_{i}q_{i}}\frac{dq_{i}}{dt}=\frac{1}{d_{j}q_{j}}\frac{dq_{j}}{dt}%
\Rightarrow
\end{equation*}%
\begin{equation*}
\int \frac{1}{d_{j}q_{j}}dq_{i}=\int \frac{1}{d_{j}q_{j}}dq_{j}+C\Rightarrow
\end{equation*}%
\begin{equation}
\frac{\ln q_{i}(t)}{d_{i}}=\frac{\ln q_{j}(t)}{d_{j}}+C.  \tag{C.2}
\label{eq:secondeq}
\end{equation}%
Considering $t=0$ in equation (\ref{eq:secondeq}) we obtain 
\begin{equation}
C=\frac{\ln q_{i}(0)}{d_{i}}-\frac{\ln q_{j}(0)}{d_{j}}.  \tag{C.3}
\label{eq:newCeq}
\end{equation}%
Combining (\ref{eq:newCeq}) with (\ref{eq:secondeq}) we obtain 
\begin{equation*}
\frac{\ln q_{i}(t)-\ln q_{i}(0)}{d_{i}}=\frac{\ln q_{j}(t)-\ln q_{j}(0)}{%
d_{j}}\Rightarrow
\end{equation*}%
\begin{equation}
\left( \frac{q_{i}(t)}{q_{i}(0)}\right) ^{1/d_{i}}=\left( \frac{q_{j}(t)}{%
q_{j}(0)}\right) ^{1/d_{j}}.  \tag{C.4}  \label{eq:thirdeq}
\end{equation}%
Equation (\ref{eq:thirdeq}) holds for any pair $i,j$ from the same $L$%
-class, so that 
\begin{equation*}
\left( \frac{q_{i}(t)}{q_{i}(0)}\right) ^{1/d_{i}}=\lambda (t)\Rightarrow
\end{equation*}%
\begin{equation*}
q_{i}(t)=q_{i}(0)\lambda (t)^{d_{i}}
\end{equation*}%
for some $L$-class specific $\lambda (t)$. It is clear from equation (\ref%
{rep}) and the fact that $L(\bar{v}(q))$ is increasing whenever there is
heterogeneity of $L$ values within the population that for the $L_{\max }$%
-class the corresponding value $\lambda (t)$ is always increasing and for
any other class it is either always decreasing, or starts by increasing and
then eventually switches to decreasing, when the population size passes the
corresponding threshold (\ref{ncritical}). Since $\lambda (t)$ is bounded
above and below, and a monotonic function (decreasing or increasing) then it
converges. Letting $\lambda =\lim_{t\rightarrow \infty }\lambda (t)$ gives%
\begin{equation}
q_{i}=\lim_{t\rightarrow \infty }q_{i}(t)=q_{i}(0)\lambda ^{d_{i}}. 
\tag{C.5}  \label{qlambda}
\end{equation}

We know that $\sum_{i}q_{i}=1$,\ thus for at least one $L$-class the
corresponding $\lambda (t)$\ should not converge to $0$. The system (\ref%
{intra}),(\ref{inter}) shows that it will be $L_{\max }$-class. However, the
above reasoning used coordinates describing the strategy frequencies in the
whole population (a metasimplex coordinates, Argasinski 2006). According to (%
\ref{opposite}),\ $q_{i}(0)$\ can be described in the coordinates of the
system (\ref{intra}) and (\ref{inter}) and after change of the indices $%
i=k_{a}^{l}$\ where $l$\ is the index of the $L$-class and $a$\ is the index
of the strategy within this $L$-class, we have $q_{k_{a}^{l}}=g_{l}q_{a}^{l}$%
. Thus the rest-point will contain only the $L$-maximizing strategies, so
that the state of the $L_{\max }$-class will be equivalent to the state of
the whole population (i.e according to (\ref{opposite}) $g_{\max }=1$\ and $%
q_{k_{i}^{\max }}=q_{i}^{\max }$), but frequencies $q_{k_{i}^{\max }}(0)$\
will not sum to 1. However, from (\ref{opposite}) we have $q_{k_{a}^{\max
}}(0)=g_{\max }(0)q_{a}^{\max }(0)$. Then (\ref{qlambda}) for the $L_{\max }$%
-class can be presented as: 
\begin{equation}
q_{a}^{\max }=q_{a}^{\max }(0)g_{\max }(0)\lambda ^{d_{a}}.
\end{equation}

\newpage

\bigskip \textbf{References}

Argasinski, K., 2006. Dynamic multipopulation and density dependent
evolutionary games related to replicator dynamics. A metasimplex concept.
Mathematical Biosciences 202, 88-114.

Argasinski, K., Broom, M., 2012. Ecological theatre and the evolutionary
game: how environmental and demographic factors determine payoffs in
evolutionary games Journal of Mathematical Biology DOI
10.1007/s00285-012-0573-2 (open access).

Argasinski, K., Koz\l owski, J., 2008. How can we model selectively neutral
density dependence in evolutionary games. Theor Pop Biol 73 250-256.

Barbault, R., 1987. Are still r-selection and K-selection operative
concepts?. Acta Oecologica-Oecologia Generalis 8: 63-70.

Brommer, J., 2000. The evolution of fitness in life-history theory, Biol.
Rev. (2000), 75, pp. 377-404.

Brommer, J., Kokko, H., 2002. Reproductive timing and individual fitness,
Ecology Letters, (2002) 5: 802-810.

Charlesworth, B., Leon, J. A., 1976. The relation of reproductive effort to
age. American Naturalist 110, 449-459.

Charnov, E.L., W.M. Schaffer. 1973. Life history consequences of natural
selection: Cole's result revisited. American Naturalist 107:791-793

Cole, L.C. 1954. The population consequences of life history phenomena. Q.
Rev. Biol., 29: 103-137

Cressman, R., 1992. The Stability Concept of Evolutionary Game Theory.
Springer.

Cressman R., Garay J., 2003\ Evolutionary stability in Lotka--Volterra
systems, J. Theor. Biol. 222 233.

Cressman R, Garay J., 2003 Stability in N-species coevolutionary systems.
Theor Pop Biol 64:519--533

Cressman R, Garay J. Hofbauer J 2001, Evolutionary stability concepts for
N-species frequency-dependent interactions. J. theor. Biol. 211 1-10.

Cressman, R., Krivan, V., and Garay, J., 2004. Ideal Free Distributions,
Evolutionary Games, and Population Dynamics in Multiple-Species
Environments. Am Nat 164, 473-489.

Cressman, R., Krivan, V., 2006. Migration Dynamics for the Ideal Free
Distribution. Am Nat 168, 384-397.

Cressman, R., Krivan, V., 2010. The ideal free distribution as an
evolutionarily stable state in density-dependent population games. Oikos,
119: 1231-1242.

Dercole, F., Rinaldi, S., 2008. Analysis of evolutionary processes.
Princeton University Press.

Dieckmann U, Law R 1996 The dynamical theory of coevolution: a derivation
from stochastic ecological processes. J Math Biol 34: 579-612],

Dieckmann, U., Metz, J.A.J., 2006. Surprising evolutionary predictions from
enhanced ecological realism. Theor Pop Biol 69, 263-281.

Gabriel, J.P., Saucy, F., Bersier, L.F., 2005. Paradoxes in the logistic
equation? Ecol. Model. 185, 147--151.

Geritz S.A.H., \'{E}. Kisdi, G. Mesz\'{e}na \& J.A.J. Metz 1998,
Evolutionarily singular strategies and the adaptive growth and branching of
the evolutionary tree. Evol Ecol 12: 35-57

Geritz, Stefan AH, and \'{E}va Kisdi. "Mathematical ecology: why mechanistic
models?." Journal of mathematical biology 65.6 (2012):

1411-1415.

Getz, W.M., 1993. "Metaphysiological and evolutionary dynamics of
populations exploiting constant and interactive resources -- r-K selection
revisited". Evolutionary Ecology 7 (3): 287-305.

Ginzburg, L.R., 1992. Evolutionary consequences of basic growth equations.
Trends Ecol. Evol. 7, 133.

Gyllenberg M.,. Metz J. A. J, Service R. 2011 When do optimisation arguments
make evolutionary sense? p. 233-268 in F. A. C. C. Chalub and J. F.
Rodrigues (eds.) The Mathematics of Darwin's Legacy, Birkhauser

Hofbauer, J., Sigmund, K., 1988. The Theory of Evolution and Dynamical
Systems. Cambridge University Press.

Hofbauer, J., Sigmund, K., 1998. Evolutionary Games and Population Dynamics.
Cambridge University Press.

Hui, C., 2006. Carrying capacity, population equilibrium, and environment's
maximal load. Ecol. Model. 192, 1--2, 317--320

Koz\l owski, J., 1980. Density dependence, the logistic equation, and r- and
K-selection: A critique and an alternative approach. Evolutionary Theory 5,
89-101.

Koz\l owski, J., 1992. Optimal allocation of resources to growth and
reproduction: Implications for age and size at maturity.. Ternds Ecol. Evol.
7: 15-19.

Koz\l owski, J., 1993. Measuring fitness in life-history studies. Trends
Ecol. Evol., 8: 84-85.

Koz\l owski, J., 1996. Optimal initial size and adult size of animals:
consequences for macroevolution and community structure. Am. Nat.
147:101-114.

Koz\l owski, J., 2006. Why life histories are diverse Pol. J. Ecol 54 4
585-604.

Kuno, E., 1991. "Some strange properties of the logistic equation defined
with r and K -- inherent defects or artifacts". Researches on Population
Ecology 33: 33-39.

Lomnicki, A., 1988. Population ecology of individuals, Princeton University
Press.

MacArthur, R H., Wilson, E.0., 1967 The theory of island biogeography.
Princeton University Press.

Maynard Smith, J. 1982 Evolution and the Theory of Games. Cambridge
University Press.

Metz, J.A.J., Diekmann, O 1986, The dynamics of physiologically structured
populations. Springer Verlag, Lecture Notes in Biomathematics, 68] chapter
IV, section 1.2.

Metz, J.A.J., Nisbet R.M. and Geritz S.A.H., 1992. How should we define
`fitness' for general ecological scenarios? TREE, 7, 6, 198-202.

Metz, J.A.J., S.A.H. Geritz, G. Mesz\'{e}na, F.J.A. Jacobs, J.S. van
Heerwaarden, 1996. Adaptive dynamics, a geometrical study of the
consequences of nearly faithful reproduction. In: Stochastic and spatial
structures of dynamical systems. S.J. van Strien \& S.M. Verduyn Lunel, eds.
1996 pp. 183-231. North-Holland,

Metz J.A.J., Mylius S.D., Diekmann, O., 2008. When does evolution optimize?
Evolutionary Ecology Research, 10: 629-654

Metz J.A.J., Mylius, S.M., Diekmann, O. 2008b Even in the odd cases when
evolution optimizes, unrelated population dynamical details may shine
through in the ESS. Evol Ecol Res 10: 655-666

Metz J.A.J.\ 2008 Fitness. Pp. 1599-1612 in S.E. J\o rgensen \& B.D. Fath
(Eds) Evolutionary Ecology. Vol. 2 of Encyclopedia of Ecology. Elsevier

Morris D.W., 2011 Adaptation and habitat selection in the eco-evolutionary
process Proc. R. Soc. B 278 1717 2401-2411

Mylius, S.D., Diekmann, O., 1995. On evolutionarily stable life histories,
optimization and the need to be specific about density dependence. Oikos,
74: 218-224.

Perrin, N., Sibly, R.M., 1993.\ Dynamic models of energy allocation and
investment. Annu. Rev. Ecol. 7: 576-592.

Pelletier F., Garant D., Hendry A.P., 2009, Eco-evolutionary dynamics\ Phil.
Trans. R. Soc. B 364, 1483-1489

Post D.M. Palkovacs E.P. 2009. Eco-evolutionary feedbacks in community and
ecosystem ecology: interactions between the ecological theatre and the
evolutionary play. Philos Trans R Soc Lond B Biol Sci. 364: 1629-40.

Roff, D.A., 1992. The Evolution of Life Histories, Theory and Analyses.
Chapman \& Hall.

Roff, D.A., 2008. Defining fitness in evolutionary models. J Gen 87, 339-348.

Schoener T.W., 2011, The Newest Synthesis: Understanding the Interplay of
Evolutionary and Ecological Dynamics. Science 331, 426

Stearns, S.C. 1977. Evolution of life-history traits -- critique of theory
and a review of data. Ann. Rev. of Ecology and Systematics 8: 145-171.

Stearns S.C. 1992. The Evolution of Life Histories. Oxford University Press.

Taylor, P.D., Williams, G.C., 1984. Demographic parameters at evolutionary
equilibrium. Can. J. Zool. 62: 2264-2271.

Verhulst, P.F. 1838. Notice sur la loi que la population pursuit dans son
accroissement. Corresp. Math. Phys. 10: 113--121.

Werner, E.E., Anholt, B.R., 1993 Ecological consequences of the trade-off
between growth and mortality rates mediated by foraging activity. Am. Nat.
142:242-272.

Zhang F., Hui C., 2011, Eco-Evolutionary Feedback and the Invasion of
Cooperation in Prisoner's Dilemma Games. PLoS ONE 6(11): \newline
e27523.doi:10.1371/journal.pone.0027523

\end{document}